\documentclass[%
 reprint,
superscriptaddress,
 amsmath,amssymb,
 aps,
]{revtex4-2}
\usepackage{graphicx}
\usepackage[export]{adjustbox}
\usepackage{graphicx}
\usepackage{dcolumn}
\usepackage{bm}
\usepackage{dsfont}
\usepackage{xcolor}
\usepackage{amsmath}
\usepackage{braket}
\usepackage{qcircuit}
\usepackage{hyperref} 

\begin{document}

\preprint{APS/123-QED}

\title{Variational Quantum Circuits for Multi-Qubit Gate Automata}

\author{Arunava Majumder}
 \email{Arunava.Majumder@uibk.ac.at}
\affiliation{%
Institute for Theoretical Physics, University of Innsbruck, Technikerstr. 21a, A-6020 Innsbruck, Austria 
}%

\author{Dylan Lewis}
\affiliation{
Department  of  Physics  and  Astronomy,  University  College  London, London  WC1E  6BT,  United  Kingdom
}

\author{Akshaya Jayashankar} 
\affiliation{%
Centre for Quantum Engineering Research and Education, TCG CREST, First floor, Bengal Eco Intelligent Park, Sector V Salt Lake, Kolkata-700091, India 
}%

\author{V. S. Prasannaa}
\affiliation{%
Centre for Quantum Engineering Research and Education, TCG CREST, First floor, Bengal Eco Intelligent Park, Sector V Salt Lake, Kolkata-700091
}%
\affiliation{Academy of Scientific and Innovative Research (AcSIR), Ghaziabad- 201002, India} 

\author{Sougato Bose}
\affiliation{
Department  of  Physics  and  Astronomy,  University  College  London, London  WC1E  6BT,  United  Kingdom
}


\date{\today}

\begin{abstract}
Implementing quantum operations in the form of natural Hamiltonian dynamics is desirable, since they almost require no external control or feedback. In this work, we propose a NISQ-friendly quantum-classical hybrid approach to designing a time-independent Hamiltonian that generates a given multi-qubit unitary. In particular, we execute a Variational Quantum Algorithm, whose ansatz is carefully chosen to be a sequence of appropriately parametrized unitaries describing at most two-qubit nearest neighbour interactions, dictating the target unitary. Subsequently, we apply our approach to simulate multi-qubit target gates,  with  and without stochastic noise. We demonstrate that our strategy allows us to  implement a Toffoli gate with sufficiently high fidelity, as compared to the other similar techniques. Our approach is an example of the usage of quantum computing for the design of quantum computers. 
\end{abstract}

\maketitle

\section{\label{sec:introduction}Introduction}

Quantum computers offer the promise of delivering faster solutions to certain hard problems than their classical counterparts~\cite{kitaevqpe,shor1999polynomial, grover1996fast,Harrow2009QuantumEquations}. The gate-based model of quantum computing serves as a very useful tool for carrying out calculations, but is prone to serious errors due to the interference of the qubits with the environment. Therefore, devising novel approaches for implementing quantum gates emerges as a timely alternative. To that end, some recent studies propose realization of multi-qubit gates through \emph{time-independent} Hamiltonians~\cite{Banchi2016,Eloie2018,Innocenti2018,innocenti2020supervised,mortimer2021evolutionary} with nearest-neighbour interactions. A network evolving under Hamiltonian dynamics can realize the action of the desired \emph{multi-qubit gate} without necessitating external control, and such a technique is often referred to as `{gate automata}'. For example, in Refs.~\cite{Banchi2016,Innocenti2018,innocenti2020supervised}, the authors demonstrate their approach of using time-independent Hamiltonians for gate designs to systematically obtain the generators of the target unitaries via a supervised learning approach. Similarly, in Ref.~\cite{Eloie2018}, a quantum arithmetic unit is realized by carefully optimizing the interaction strengths of the underlying dynamics, using differential evolution. A more recent work~\cite{lewis} obtains the strengths of the Hamiltonian using a geodesic algorithm.
Such a technique of using a time-independent Hamiltonian to realize quantum gates can be thought of as a \emph{single-shot} strategy where the interaction is simply switched `on' once and then switched `off', which would allow one to accomplish a target unitary gate. This approach is in contrast to some existing strategies that directly implement higher-order quantum gates, such as the Toffoli gate, via a quantum control route~\cite{d2007introduction,dong2010quantum,spiteri2018quantum}, which requires the interactions to be frequently modulated. 

In our work, we devise an attractive Noisy Intermediate-Scale Quantum (NISQ)-friendly approach to designing multi-qubit gate automata, via the quantum-classical hybrid variational quantum algorithms (VQAs). In recent times, VQAs~\cite{cerezo2021variational, bharti2022noisy} have emerged as potential candidates to provide a quantum advantage in the NISQ era~\cite{preskill2018quantum}. VQAs can be employed to solve a variety of optimisation~\cite{moll2018quantum} and linear algebra problems~\cite{VQLSquantum}, as well as a prospect to simulate quantum chemical systems~\cite{peruzzo2014variational,mcclean2016theory,peng2020simulating,lyu2020accelerated,tilly2021variational,huang2022robust,VQEFedorov}. In the current work, we use a VQA to optimize the values of the couplings that are required in realizing a given target multi-qubit quantum gate via a time-independent Hamiltonian. The quantum part of the procedure involves evaluating a parametrized quantum circuit for a given set of parameters in each iteration. The parameters are updated for the next iteration via a cost function optimization routine on a classical computer. The optimized parameters are the interaction strengths of the time-independent Hamiltonian that generate the target unitary. Further, we demonstrate the applicability of our approach to designing Toffoli and Parity gates. 

We organize the rest of the manuscript as follows--- Sec.~\ref{sec:theory} describes our VQA scheme, along with details on the functional form of the time-independent Hamiltonian that we consider. We present the optimized results for the Toffoli gate and the Parity gate in Sec.~\ref{sec:results}. We conclude in Sec.~\ref{sec:conclusion}. 

\section{\label{sec:theory}Theory and Methodology} 

 
\subsection{The parametrized unitary for VQA}\label{sec:A}
\begin{figure}[t]
    \centering
    \includegraphics[scale=0.2]{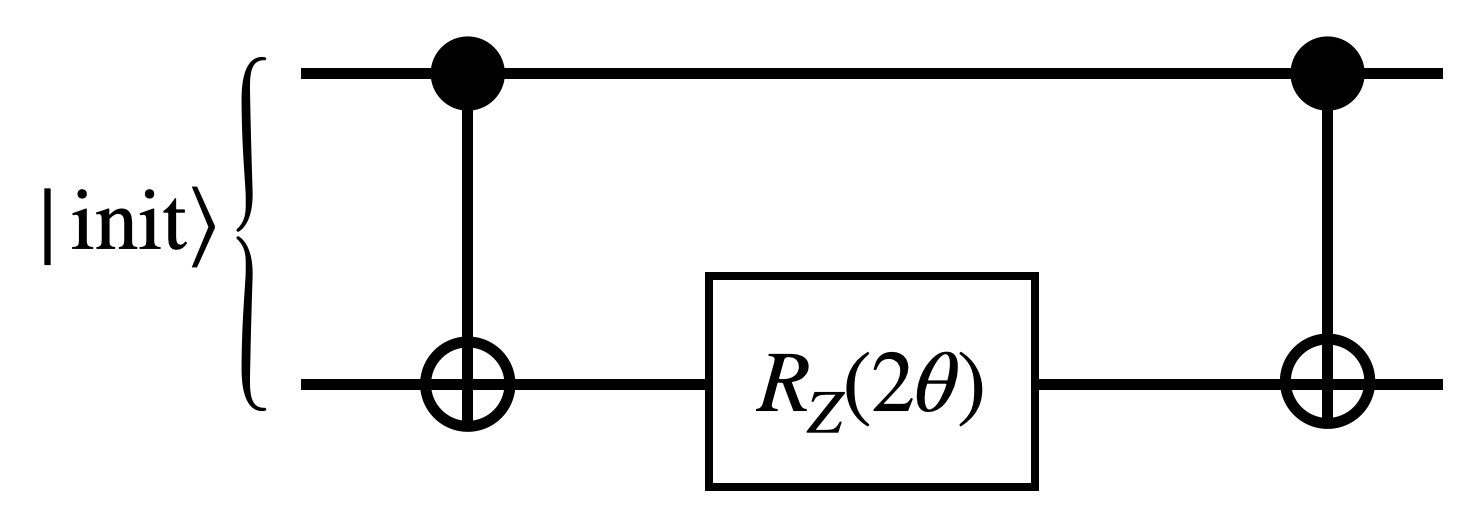}
    \caption{Circuit to implement the evolution of each term in the 2-qubit Ising interaction of the form $e^{i \theta \sigma_Z \otimes \sigma_Z}$, where $\theta$ is chosen as $J_{ij}$ for a given $i$ and $j$ as shown in Eq.~\ref{eq:ising}. The input state is denoted as $\ket{\mathrm{init}}$ in the figure. The $e^{i \theta \sigma_{X} \otimes\sigma_{X}}$ and $e^{i \theta \sigma_{Y} \otimes \sigma_{Y}}$ interactions can be similarly implemented with appropriate basis transformations at the beginning and end of the circuit. }
    \label{fig:ising_interaction_circuit}
\end{figure}

In our work, we aim to simulate an $n$-qubit gate using  Hamiltonian interaction on a bunch of $n$ qubits. We consider a two-local Hamiltonian given by 

\begin{eqnarray}
\label{eq:general_hamiltonian}
H(\bm{\theta}) &=& \sum_i h_i^{\alpha} \sigma_i^{\alpha}+\sum_{i,j} J_{ij}^{\alpha \beta} \sigma_i^{\alpha}\sigma_j^{\beta} \nonumber \\  
&\equiv& \sum_{K} H_{K}(\theta_{K}), 
\end{eqnarray} 
\noindent where $H_{K}(\theta_{K})$= $h_i^{\alpha} \sigma_i^{\alpha}+ J_{ij}^{\alpha \beta} \sigma_i^{\alpha}\sigma_j^{\beta}$ and  $\{\theta_{K}\}$=$\{h_i^{\alpha}, J_{ij}^{\alpha \beta}\} $ are the parameters that one has to optimize via a VQA, in order to implement the desired multi-qubit gate. $\sigma^\alpha$, $\sigma^\beta$ $\in$ $\{\sigma^X,\sigma^Y,\sigma^Z\}$, are the Pauli operators and $\{i,j\}$ denote the qubit index. For our requirement of a \emph{physical} gate set, we require $\alpha = \beta$, such that any two spins interact through $XYZ$ Heisenberg interactions. Thus, our unitary generated using such a Hamiltonian is 

\begin{eqnarray}
\label{eq:U_theta_with_error_term}
U_{QC}(\bm{\theta}) = e^{-i H(\bm{\theta})} \approx \Bigg(\prod_{K=1}^{Q}e^{-i H_K(\theta_K) / m }\Bigg)^m, 
\end{eqnarray}

\noindent where we have used first-order Trotterization with $m$ Trotter steps. Thus, in our VQA, the quantum circuit for $U_{QC}(\bm{\theta})$ built out of $m$ repetitions of each of the product terms in Eq.~\ref{eq:U_theta_with_error_term}. For example, consider the case of an Ising interaction, with the Hamiltonian 
\begin{equation} \label{eq:ising}
 H = \sum_{ij} J_{ij}^{ZZ} \sigma_Z^i \otimes \sigma_Z^j,  
\end{equation} 
where, $J_{ij}^{ZZ}$ are coupling strengths between sites $i$ and $j$. The evolution in this case could be implemented as a sequence of circuits with each individual circuit  described as in Fig.~\ref{fig:ising_interaction_circuit}. 

It is worth noting at this point that the choice of ansatz in a VQA is crucial, as it must be able to efficiently encode the solution to the problem of interest. The specific structure of the ansatz generally depends on the task at hand, where the details of the problem itself are used to tailor the ansatz~\cite{mcardle2019variational}. In general, there can be several ways to design a suitable problem-inspired ansatz, for example, quantum approximate optimization algorithm~\cite{farhi2014quantum}, hardware efficient ansatze~\cite{kandala2017hardware}, quantum-optimal-control-inspired ansatz~\cite{choquette2021quantum}, symmetry motivated ansatze~\cite{lyu2022symmetry}, and Hamiltonian variational ansatz (HVA)~\cite{wecker2015progress,wiersema2020exploring}. We have chosen an HVA-inspired ansatz for our work, and the corresponding unitary is given by Eq. \ref{eq:U_theta_with_error_term}. We also note that $t$ is set to $1$ in our unitary. 

\subsection{The cost function for VQA}\label{sec:B} 

We obtain the optimal values of a Hamiltonian evolution that mimics the target unitary using the following cost function: 

\begin{eqnarray}
    \label{eq:cost_function}
    C(\bm{\theta}) &=& 1- F(\bm{\theta}) \nonumber \\
    &=&1 - \frac{1}{2^{2n}}\Big\vert \textrm{Tr}\left[U_\textrm{target}^\dagger U_{\textrm{QC}}(\bm{\theta})\right]\Big\vert^2, 
\end{eqnarray}
\noindent where we recall that $n$ refers to the number of qubits on which the unitary acts. $U_{\textrm{QC}}(\bm{\theta})$ is the parametrized unitary circuit obtained as described in Eq.~\ref{eq:U_theta_with_error_term}.  $U_{\textrm{target}}$ is the target unitary and the term $F(\bm{\theta})$=$\mid \textrm{Tr}[U_\textrm{target}^\dagger U_{QC}(\bm{\theta})]\mid^2/2^{2n}$ is referred to as the operator fidelity measure. Minimizing the cost function, $C(\bm{\theta})$, described by Eq.~\ref{eq:cost_function} results in a set of optimal values, $\theta_{\rm opt}$, which then allows us to realize the target unitary $U_{\textrm{target}}$ via $U_{\textrm{QC}}(\bm{\theta_{\rm opt}})$. 

\subsection{Quantum circuit for the cost function and VQA procedure}

\begin{figure*}[t]
\includegraphics[scale=0.32]{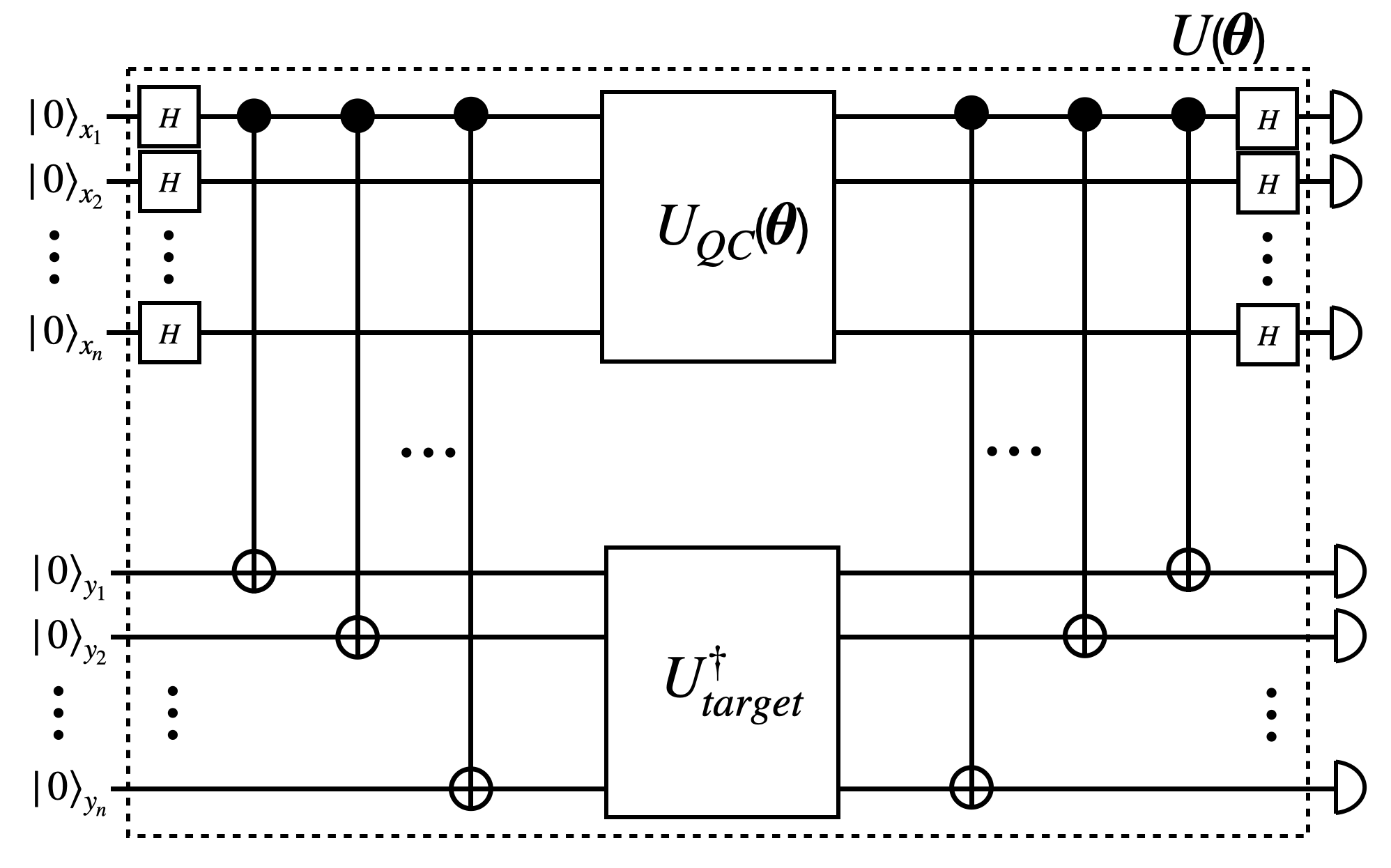}
\caption{The quantum circuit for Hilbert-Schmidt test. Two unitaries $U_{QC}(\bm{\theta})$ and $U^{\dagger}_{target}$ are acting on $n$ qubit basis states. Measuring all the $2n$ qubits in $|0\rangle$ state will give the output $\bigg\vert \textrm{Tr}\left[U_\textrm{target}^\dagger U_{\textrm{QC}}(\bm{\theta})\right]\bigg\vert^2/2^{2n}$. }
\label{fig:hilbert_schmidt_test_circuit}
\end{figure*}

\begin{figure*}[t] 
\centering
\includegraphics[scale=0.36]{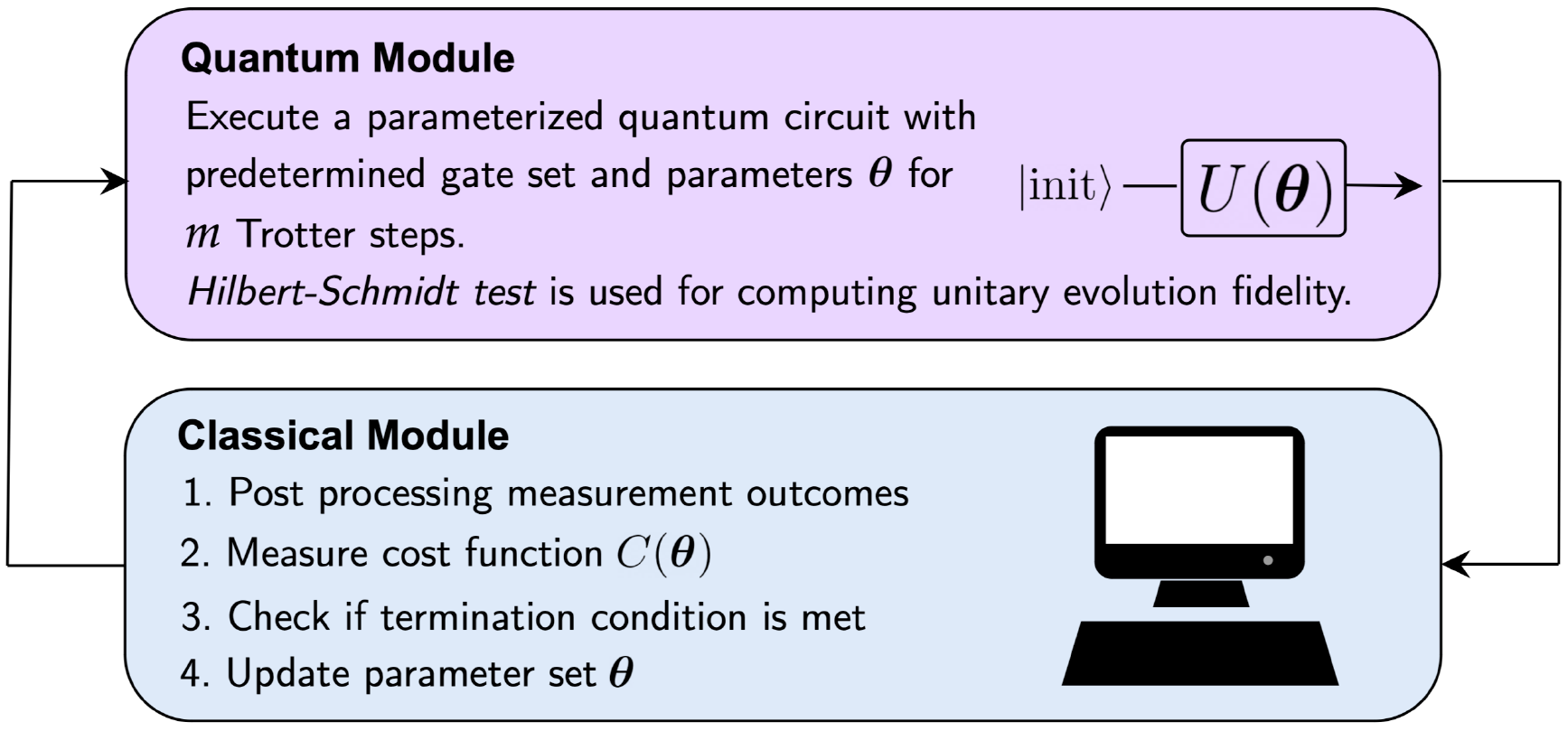}
\centering \caption{Schematic illustrating the working of our variational quantum algorithm. }
\label{fig:vqa}
\end{figure*}

We now piece together the information from the preceding two subsections~\ref{sec:A} and \ref{sec:B}, and combine them with the fact that our cost function can be represented as the Hilbert-Schmidt test quantum circuit~\cite{khatri2019quantum} as shown in Fig.~\ref{fig:hilbert_schmidt_test_circuit}. As the circuit diagram indicates, we require $2n$ qubits, such that the evolution operator $U_\textrm{QC}(\bm{\theta})$ acts on the first $n$ qubits, while on the next $n$ qubits, the target unitary gate, $U_\textrm{target}$, is applied. The two subsystems on which both unitaries act are first maximally entangled (which requires $2n$ gates). After both sets of gates are applied to each subsystem separately, a global Bell state measurement (requiring another $2n$ gates) is performed to obtain the overall fidelity.
Following this step, an updated set of parameters is generated using a classical optimizer routine. For this purpose, we employ the Adagrad optimizer ~\cite{duchi2011adaptive}. This updated set is used in computing the cost function for the next iteration, following which the parameters are again updated. The process is repeated until the cost $C(\bm{\theta})$ is sufficiently small. Thus, we obtain variationally the approximate evolution of $H$ mimicking the target quantum gate. We used the Pennylane software~\cite{bergholm2018pennylane} for all our simulations. Fig.~\ref{fig:vqa} illustrates the schematic of our VQA. 

\section{\label{sec:results}Results}

We first demonstrate the ability of VQA to find an optimal time-independent Hamiltonian with single and two-qubit interaction terms that implement the a) Toffoli gate b) Parity gate. This will be followed by a discussion on VQA for stochastic Hamiltonians. 

\begin{figure}[!ht]
 \includegraphics[scale=0.67]{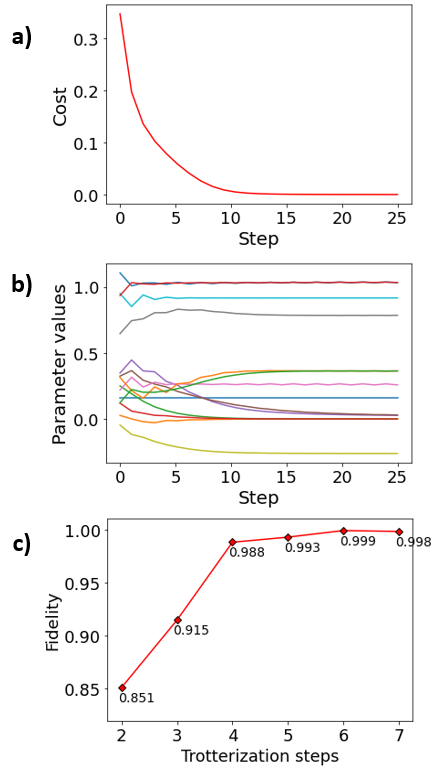}
 \centering 
 \caption{
Sub-figure (a) shows training data for cost optimization with Trotter step $m=6$, and with gradient descent optimizer. In the plot, `Step' refers to iteration number. Sub-figure (b) presents the evolution of the parameters with the optimization steps. The plot shows that all of the parameters saturate after a few steps. Sub-figure (c) shows the improvement of fidelity between $U_\textrm{Toff}$ and $e^{-iH(\bm{\theta_\textrm{opt}})}$, for $t=1$, as a function of the number of Trotter steps ($m$), where $\bm{\theta_\textrm{opt}}$ is the set of optimal parameters in the parameterized Hamiltonian of Eq.~\eqref{eq:general_hamiltonian} for each number of Trotterization steps. }
\label{fig:fidelity_vs_trotterization_steps}
\label{fig:Toffoli results}
\end{figure}

\begin{figure}[h]
 \centering
 \includegraphics[scale=0.73]{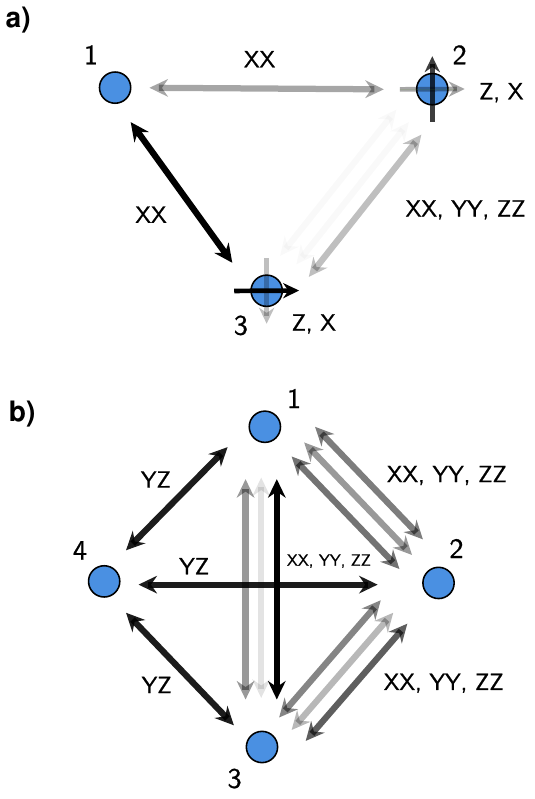}
 \centering 
\caption{Interaction diagrams with single and two-qubit interactions for a) Toffoli gate and b) Parity gate. All the parameters are found from variational circuit optimization and give fidelities of $F > 0.99$. The Trotter steps used are $m=6$ for (a) and $m=5$ for (b). The interaction strengths for (a): $J^{X}_{1}=1.09$, $J^{Z}_{1}=2.35$, $J^{X}_{2}=3.11$, $J^{Z}_{2}=-0.78$, $J^{XX}_{12}=0.07$, $J^{YY}_{12}=0.07$, $J^{ZZ}_{12}=0.78$, $J^{XX}_{13}=1.089$, $J^{ZZ}_{23}=3.11$. Interaction strengths for (b): $J^{XX}_{12}=1.42$, $J^{YY}_{12}=1.04$, $J^{ZZ}_{12}=1.30$, $J^{XX}_{23}=1.23$, $J^{YY}_{23}=0.73$, 
$J^{ZZ}_{23}=1.60$, 
$J^{XX}_{13}=1.03$, 
$J^{YY}_{13}=0.29$, 
$J^{ZZ}_{13}=2.57$, 
$J^{ZY}_{24}=2.37$, 
$J^{ZY}_{14}=2.29$, 
$J^{ZY}_{34}=2.30$.}
 \label{fig:toffoli_and_parity_gate_interactions}
\end{figure}

\textbf{Toffoli gate:} We observe from Fig.~\ref{fig:Toffoli results} that at $m=6$, the unitary evolution quickly finds a minimum parameter set $\bm{\theta}_\textrm{min}$, such that the cost function, $C(\bm{\theta_{\min}})$ is close to 0. The figure also shows that remarkably, the operator fidelity, $F(\bm{\theta}_\textrm{min})$, is greater than 0.99 even for low values of $m$. The optimal interaction strengths thus obtained via our search procedure  for the Toffoli gate are presented in Fig.~\ref{fig:toffoli_and_parity_gate_interactions}(a). In this figure, the edge colors are varied according to their interacting strength $J_{ij}^{\sigma_{n_1}\sigma_{n_2}}$. The black edges correspond to strong coupling interaction and the faded ones describe weaker interactions.

\textbf{Parity gate:} An extension of the variational quantum optimization algorithm involves exploring if a parity check procedure (to detect possible errors in the surface code) can be implemented directly by a time-independent Hamiltonian. Assume that one needs to obtain the parity, $p$, of three qubits, described in the computational basis  $z_i \in \{0,1\}$ with $i=1,2,3$. The parity $p=0$ when there is either no $|1\rangle$ or even number of $|1\rangle$s, and $p=1$ otherwise. This is achieved by implementing the interaction $P = \sigma^Z_1 \otimes \sigma^Z_2 \otimes \sigma^Z_3 \otimes \sigma^Y_4$ at $t=\pi/4$ measuring the parity $p = z_1 \oplus z_2 \oplus z_3$ as shown below,
\begin{align}
    U(\tfrac{\pi}{4})|z_1 z_2 z_3\rangle |0\rangle &= \frac{1}{\sqrt{2}}\left(|z_1 z_2 z_3\rangle |0\rangle - i P |z_1 z_2 z_3\rangle |0\rangle \right) \nonumber \\ 
    &= \frac{1}{\sqrt{2}}|z_1 z_2 z_3\rangle \left( |0\rangle - (-1)^p |1\rangle \right),
\end{align}
where $U(\frac{\pi}{4}) = e^{-iP \frac{\pi}{4}}$. The state of the last qubit is either $|+\rangle$ or $|-\rangle$, depending on whether the parity $p$ is even or odd respectively. one could then perform an $x$-basis measurement by applying a Hadamard to the last qubit followed by a computational basis measurement. The measurement outcome of the final qubit thus corresponds to the parity of the first three qubits. Fig.~\ref{fig:toffoli_and_parity_gate_interactions}(b) shows the single-qubit and two-qubit interactions required to perform the three-qubit parity check evolution with a fidelity of more than 0.99. We note that in the case of sparse multi-qubit unitaries, the number of parameters that one may have to search for would only grow polynomially with the sparsity of the unitary itself, based on Ref.~\cite{berry}.

\begin{figure}[h]
 \centering
 \includegraphics[scale=0.45]{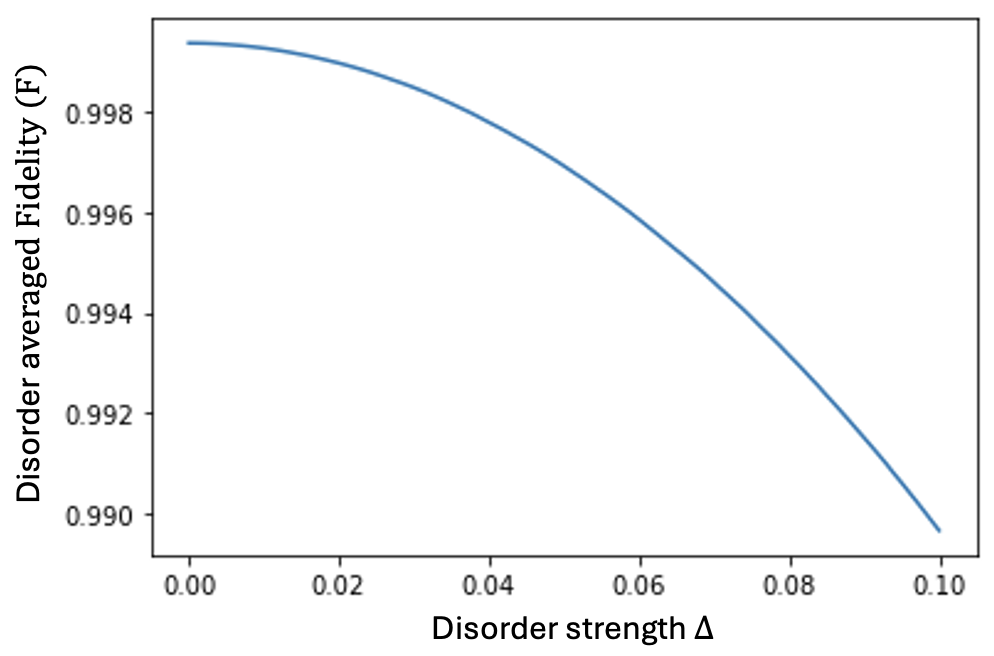}
 \centering 
\caption{Figure presenting our results for disorder-averaged fidelity, F, versus disorder strength, $\Delta$. }
 \label{fig:disordered_hamiltonian}
\end{figure}

\noindent \textbf{VQA for stochastic Hamiltonians}:\label{sec:stochastic}
We now demonstrate the performance of VQA assuming that the couplings are disordered in the following way.
\begin{equation}
    \label{eq:general_hamiltonian}
    H = \sum_{i,j} J_{ij}^{\alpha \beta; \delta} \sigma_i^{\alpha}\sigma_j^{\beta},
\end{equation}
where the disordered couplings are drawn from a normal distribution and is given as  $J_{ij}^{\alpha \beta; \delta}=\Bar{J}+\delta$, such that $\delta$ $\in$ $[-\Delta,\ \Delta]$, with $\Delta$ being the disorder strength. We plot disorder-averaged fidelity versus disorder strength in Fig.~\ref{fig:disordered_hamiltonian} for the case of the Toffoli gate, demonstrating that this gate is still robust even in the presence of disorder. 

\section{Conclusion}\label{sec:conclusion}

In this work, we demonstrate that the NISQ-friendly variational quantum algorithm can find the coupling parameters for multi-qubit quantum gates with high fidelity. Besides applying our approach to the Toffoli gate, we also obtain the optimal coupling strengths for the multi-qubit Parity gate, in terms of  two-qubit interactions in Section~\ref{sec:results}. We comment here that the latter could find use in quantum error correction~\cite{bultink2020protecting,roffe2018protecting}. Furthermore, we find that the Hamiltonian obtained via VQA is robust even in the presence of disorder. 

Typically, a variational quantum algorithm often suffers from barren plateaus \cite{mcclean2018barren,cerezo2021cost} as we increase the number of qubits for a problem. This can be averted by using the local Hilbert-Schmidt test, as described in Ref. \cite{khatri2019quantum}. Indeed, we observe that our problem does not suffer from the barren plateaus issue. Additionally, our ansatz is problem-inspired, and we impose certain conditions to our Hamiltonian so that we are able to reduce the variational parameters we start with. We intuit that the  use of inductive techniques may provide further resolution -- optimal values of couplings derived for gates in a calculation involving fewer qubits to formulate the initial ansatz for a system with a larger number of qubits. Our results demonstrate the possibility of implementing Hamiltonian evolution to mimic multi-qubit gates, thereby circumventing the need to apply gates that are noisy throughout the computation. We thus anticipate that our work could potentially pave way for newer approaches to quantum computing so that current challenges involved in performing gate-based computation are averted. 

\section{Acknowledgements} AM acknowledges the support of the European Union (ERC Advanced Grant, QuantAI, No. 101055129) for the doctoral program in UIBK. DL acknowledges support from the EPSRC Centre for Doctoral Training in Delivering Quantum Technologies, grant ref.~EP/S021582/1. 

\bibliography{references}

\begin{thebibliography}{40}%
\makeatletter
\providecommand \@ifxundefined [1]{%
 \@ifx{#1\undefined}
}%
\providecommand \@ifnum [1]{%
 \ifnum #1\expandafter \@firstoftwo
 \else \expandafter \@secondoftwo
 \fi
}%
\providecommand \@ifx [1]{%
 \ifx #1\expandafter \@firstoftwo
 \else \expandafter \@secondoftwo
 \fi
}%
\providecommand \natexlab [1]{#1}%
\providecommand \enquote  [1]{``#1''}%
\providecommand \bibnamefont  [1]{#1}%
\providecommand \bibfnamefont [1]{#1}%
\providecommand \citenamefont [1]{#1}%
\providecommand \href@noop [0]{\@secondoftwo}%
\providecommand \href [0]{\begingroup \@sanitize@url \@href}%
\providecommand \@href[1]{\@@startlink{#1}\@@href}%
\providecommand \@@href[1]{\endgroup#1\@@endlink}%
\providecommand \@sanitize@url [0]{\catcode `\\12\catcode `\$12\catcode `\&12\catcode `\#12\catcode `\^12\catcode `\_12\catcode `\%12\relax}%
\providecommand \@@startlink[1]{}%
\providecommand \@@endlink[0]{}%
\providecommand \url  [0]{\begingroup\@sanitize@url \@url }%
\providecommand \@url [1]{\endgroup\@href {#1}{\urlprefix }}%
\providecommand \urlprefix  [0]{URL }%
\providecommand \Eprint [0]{\href }%
\providecommand \doibase [0]{https://doi.org/}%
\providecommand \selectlanguage [0]{\@gobble}%
\providecommand \bibinfo  [0]{\@secondoftwo}%
\providecommand \bibfield  [0]{\@secondoftwo}%
\providecommand \translation [1]{[#1]}%
\providecommand \BibitemOpen [0]{}%
\providecommand \bibitemStop [0]{}%
\providecommand \bibitemNoStop [0]{.\EOS\space}%
\providecommand \EOS [0]{\spacefactor3000\relax}%
\providecommand \BibitemShut  [1]{\csname bibitem#1\endcsname}%
\let\auto@bib@innerbib\@empty
\bibitem [{\citenamefont {Kitaev}(1995)}]{kitaevqpe}%
  \BibitemOpen
  \bibfield  {author} {\bibinfo {author} {\bibfnamefont {A.~Y.}\ \bibnamefont {Kitaev}},\ }\bibfield  {title} {\bibinfo {title} {Quantum measurements and the {A}belian stabilizer problem},\ }\href@noop {} {\bibfield  {journal} {\bibinfo  {journal} {arXiv preprint arXiv:9511026}\ } (\bibinfo {year} {1995})}\BibitemShut {NoStop}%
\bibitem [{\citenamefont {Shor}(1999)}]{shor1999polynomial}%
  \BibitemOpen
  \bibfield  {author} {\bibinfo {author} {\bibfnamefont {P.~W.}\ \bibnamefont {Shor}},\ }\bibfield  {title} {\bibinfo {title} {Polynomial-time algorithms for prime factorization and discrete logarithms on a quantum computer},\ }\href@noop {} {\bibfield  {journal} {\bibinfo  {journal} {SIAM review}\ }\textbf {\bibinfo {volume} {41}},\ \bibinfo {pages} {303} (\bibinfo {year} {1999})}\BibitemShut {NoStop}%
\bibitem [{\citenamefont {Grover}(1996)}]{grover1996fast}%
  \BibitemOpen
  \bibfield  {author} {\bibinfo {author} {\bibfnamefont {L.~K.}\ \bibnamefont {Grover}},\ }\bibfield  {title} {\bibinfo {title} {A fast quantum mechanical algorithm for database search},\ }in\ \href@noop {} {\emph {\bibinfo {booktitle} {Proceedings of the twenty-eighth annual ACM symposium on Theory of computing}}}\ (\bibinfo {year} {1996})\ pp.\ \bibinfo {pages} {212--219}\BibitemShut {NoStop}%
\bibitem [{\citenamefont {Harrow}\ \emph {et~al.}(2009)\citenamefont {Harrow}, \citenamefont {Hassidim},\ and\ \citenamefont {Lloyd}}]{Harrow2009QuantumEquations}%
  \BibitemOpen
  \bibfield  {author} {\bibinfo {author} {\bibfnamefont {A.~W.}\ \bibnamefont {Harrow}}, \bibinfo {author} {\bibfnamefont {A.}~\bibnamefont {Hassidim}},\ and\ \bibinfo {author} {\bibfnamefont {S.}~\bibnamefont {Lloyd}},\ }\bibfield  {title} {\bibinfo {title} {{Quantum algorithm for linear systems of equations}},\ }\href {https://doi.org/10.1103/PhysRevLett.103.150502} {\bibfield  {journal} {\bibinfo  {journal} {Physical Review Letters}\ }\textbf {\bibinfo {volume} {103}},\ \bibinfo {pages} {150502} (\bibinfo {year} {2009})}\BibitemShut {NoStop}%
\bibitem [{\citenamefont {Banchi}\ \emph {et~al.}(2016)\citenamefont {Banchi}, \citenamefont {Pancotti},\ and\ \citenamefont {Bose}}]{Banchi2016}%
  \BibitemOpen
  \bibfield  {author} {\bibinfo {author} {\bibfnamefont {L.}~\bibnamefont {Banchi}}, \bibinfo {author} {\bibfnamefont {N.~A.}\ \bibnamefont {Pancotti}},\ and\ \bibinfo {author} {\bibfnamefont {S.}~\bibnamefont {Bose}},\ }\bibfield  {title} {\bibinfo {title} {Quantum gate learning in qubit networks: {Toffoli} gate without time-dependent control},\ }\href {https://doi.org/10.1038/npjqi.2016.19} {\bibfield  {journal} {\bibinfo  {journal} {npj Quantum Information}\ }\textbf {\bibinfo {volume} {2}},\ \bibinfo {pages} {1} (\bibinfo {year} {2016})}\BibitemShut {NoStop}%
\bibitem [{\citenamefont {Eloie}\ \emph {et~al.}(2018)\citenamefont {Eloie}, \citenamefont {Banchi},\ and\ \citenamefont {Bose}}]{Eloie2018}%
  \BibitemOpen
  \bibfield  {author} {\bibinfo {author} {\bibfnamefont {L.}~\bibnamefont {Eloie}}, \bibinfo {author} {\bibfnamefont {L.}~\bibnamefont {Banchi}},\ and\ \bibinfo {author} {\bibfnamefont {S.}~\bibnamefont {Bose}},\ }\bibfield  {title} {\bibinfo {title} {Quantum arithmetics via computation with minimized external control: {The} half-adder},\ }\href {https://doi.org/10.1103/PhysRevA.97.062321} {\bibfield  {journal} {\bibinfo  {journal} {Physical Review A}\ }\textbf {\bibinfo {volume} {97}},\ \bibinfo {pages} {062321} (\bibinfo {year} {2018})}\BibitemShut {NoStop}%
\bibitem [{\citenamefont {Innocenti}\ \emph {et~al.}(2018)\citenamefont {Innocenti}, \citenamefont {Banchi}, \citenamefont {Bose}, \citenamefont {Ferraro},\ and\ \citenamefont {Paternostro}}]{Innocenti2018}%
  \BibitemOpen
  \bibfield  {author} {\bibinfo {author} {\bibfnamefont {L.}~\bibnamefont {Innocenti}}, \bibinfo {author} {\bibfnamefont {L.}~\bibnamefont {Banchi}}, \bibinfo {author} {\bibfnamefont {S.}~\bibnamefont {Bose}}, \bibinfo {author} {\bibfnamefont {A.}~\bibnamefont {Ferraro}},\ and\ \bibinfo {author} {\bibfnamefont {M.}~\bibnamefont {Paternostro}},\ }\bibfield  {title} {\bibinfo {title} {Approximate supervised learning of quantum gates via ancillary qubits},\ }\href@noop {} {\bibfield  {journal} {\bibinfo  {journal} {International Journal of Quantum Information}\ }\textbf {\bibinfo {volume} {16}},\ \bibinfo {pages} {1840004} (\bibinfo {year} {2018})}\BibitemShut {NoStop}%
\bibitem [{\citenamefont {Innocenti}\ \emph {et~al.}(2020)\citenamefont {Innocenti}, \citenamefont {Banchi}, \citenamefont {Ferraro}, \citenamefont {Bose},\ and\ \citenamefont {Paternostro}}]{innocenti2020supervised}%
  \BibitemOpen
  \bibfield  {author} {\bibinfo {author} {\bibfnamefont {L.}~\bibnamefont {Innocenti}}, \bibinfo {author} {\bibfnamefont {L.}~\bibnamefont {Banchi}}, \bibinfo {author} {\bibfnamefont {A.}~\bibnamefont {Ferraro}}, \bibinfo {author} {\bibfnamefont {S.}~\bibnamefont {Bose}},\ and\ \bibinfo {author} {\bibfnamefont {M.}~\bibnamefont {Paternostro}},\ }\bibfield  {title} {\bibinfo {title} {Supervised learning of time-independent hamiltonians for gate design},\ }\href@noop {} {\bibfield  {journal} {\bibinfo  {journal} {New Journal of Physics}\ }\textbf {\bibinfo {volume} {22}},\ \bibinfo {pages} {065001} (\bibinfo {year} {2020})}\BibitemShut {NoStop}%
\bibitem [{\citenamefont {Mortimer}\ \emph {et~al.}(2021)\citenamefont {Mortimer}, \citenamefont {Estarellas}, \citenamefont {Spiller},\ and\ \citenamefont {D'Amico}}]{mortimer2021evolutionary}%
  \BibitemOpen
  \bibfield  {author} {\bibinfo {author} {\bibfnamefont {L.}~\bibnamefont {Mortimer}}, \bibinfo {author} {\bibfnamefont {M.~P.}\ \bibnamefont {Estarellas}}, \bibinfo {author} {\bibfnamefont {T.~P.}\ \bibnamefont {Spiller}},\ and\ \bibinfo {author} {\bibfnamefont {I.}~\bibnamefont {D'Amico}},\ }\bibfield  {title} {\bibinfo {title} {Evolutionary computation for adaptive quantum device design},\ }\href@noop {} {\bibfield  {journal} {\bibinfo  {journal} {Advanced Quantum Technologies}\ }\textbf {\bibinfo {volume} {4}},\ \bibinfo {pages} {2100013} (\bibinfo {year} {2021})}\BibitemShut {NoStop}%
\bibitem [{\citenamefont {Lewis}\ \emph {et~al.}(2024)\citenamefont {Lewis}, \citenamefont {Wiersema}, \citenamefont {Carrasquilla},\ and\ \citenamefont {Bose}}]{lewis}%
  \BibitemOpen
  \bibfield  {author} {\bibinfo {author} {\bibfnamefont {D.}~\bibnamefont {Lewis}}, \bibinfo {author} {\bibfnamefont {R.}~\bibnamefont {Wiersema}}, \bibinfo {author} {\bibfnamefont {J.}~\bibnamefont {Carrasquilla}},\ and\ \bibinfo {author} {\bibfnamefont {S.}~\bibnamefont {Bose}},\ }\href@noop {} {\bibinfo {title} {Geodesic algorithm for unitary gate design with time-independent hamiltonians}} (\bibinfo {year} {2024}),\ \Eprint {https://arxiv.org/abs/2401.05973} {arXiv:2401.05973 [quant-ph]} \BibitemShut {NoStop}%
\bibitem [{\citenamefont {D’Alessandro}(2007)}]{d2007introduction}%
  \BibitemOpen
  \bibfield  {author} {\bibinfo {author} {\bibfnamefont {D.}~\bibnamefont {D’Alessandro}},\ }\href@noop {} {\bibinfo {title} {Introduction to quantum control and dynamics. hoboken}} (\bibinfo {year} {2007})\BibitemShut {NoStop}%
\bibitem [{\citenamefont {Dong}\ and\ \citenamefont {Petersen}(2010)}]{dong2010quantum}%
  \BibitemOpen
  \bibfield  {author} {\bibinfo {author} {\bibfnamefont {D.}~\bibnamefont {Dong}}\ and\ \bibinfo {author} {\bibfnamefont {I.~R.}\ \bibnamefont {Petersen}},\ }\bibfield  {title} {\bibinfo {title} {Quantum control theory and applications: a survey},\ }\href@noop {} {\bibfield  {journal} {\bibinfo  {journal} {IET Control Theory \& Applications}\ }\textbf {\bibinfo {volume} {4}},\ \bibinfo {pages} {2651} (\bibinfo {year} {2010})}\BibitemShut {NoStop}%
\bibitem [{\citenamefont {Spiteri}\ \emph {et~al.}(2018)\citenamefont {Spiteri}, \citenamefont {Schmidt}, \citenamefont {Ghosh}, \citenamefont {Zahedinejad},\ and\ \citenamefont {Sanders}}]{spiteri2018quantum}%
  \BibitemOpen
  \bibfield  {author} {\bibinfo {author} {\bibfnamefont {R.~J.}\ \bibnamefont {Spiteri}}, \bibinfo {author} {\bibfnamefont {M.}~\bibnamefont {Schmidt}}, \bibinfo {author} {\bibfnamefont {J.}~\bibnamefont {Ghosh}}, \bibinfo {author} {\bibfnamefont {E.}~\bibnamefont {Zahedinejad}},\ and\ \bibinfo {author} {\bibfnamefont {B.~C.}\ \bibnamefont {Sanders}},\ }\bibfield  {title} {\bibinfo {title} {Quantum control for high-fidelity multi-qubit gates},\ }\href@noop {} {\bibfield  {journal} {\bibinfo  {journal} {New Journal of Physics}\ }\textbf {\bibinfo {volume} {20}},\ \bibinfo {pages} {113009} (\bibinfo {year} {2018})}\BibitemShut {NoStop}%
\bibitem [{\citenamefont {Cerezo}\ \emph {et~al.}(2021{\natexlab{a}})\citenamefont {Cerezo}, \citenamefont {Arrasmith}, \citenamefont {Babbush}, \citenamefont {Benjamin}, \citenamefont {Endo}, \citenamefont {Fujii}, \citenamefont {McClean}, \citenamefont {Mitarai}, \citenamefont {Yuan}, \citenamefont {Cincio} \emph {et~al.}}]{cerezo2021variational}%
  \BibitemOpen
  \bibfield  {author} {\bibinfo {author} {\bibfnamefont {M.}~\bibnamefont {Cerezo}}, \bibinfo {author} {\bibfnamefont {A.}~\bibnamefont {Arrasmith}}, \bibinfo {author} {\bibfnamefont {R.}~\bibnamefont {Babbush}}, \bibinfo {author} {\bibfnamefont {S.~C.}\ \bibnamefont {Benjamin}}, \bibinfo {author} {\bibfnamefont {S.}~\bibnamefont {Endo}}, \bibinfo {author} {\bibfnamefont {K.}~\bibnamefont {Fujii}}, \bibinfo {author} {\bibfnamefont {J.~R.}\ \bibnamefont {McClean}}, \bibinfo {author} {\bibfnamefont {K.}~\bibnamefont {Mitarai}}, \bibinfo {author} {\bibfnamefont {X.}~\bibnamefont {Yuan}}, \bibinfo {author} {\bibfnamefont {L.}~\bibnamefont {Cincio}}, \emph {et~al.},\ }\bibfield  {title} {\bibinfo {title} {Variational quantum algorithms},\ }\href@noop {} {\bibfield  {journal} {\bibinfo  {journal} {Nature Reviews Physics}\ }\textbf {\bibinfo {volume} {3}},\ \bibinfo {pages} {625} (\bibinfo {year} {2021}{\natexlab{a}})}\BibitemShut {NoStop}%
\bibitem [{\citenamefont {Bharti}\ \emph {et~al.}(2022)\citenamefont {Bharti}, \citenamefont {Cervera-Lierta}, \citenamefont {Kyaw}, \citenamefont {Haug}, \citenamefont {Alperin-Lea}, \citenamefont {Anand}, \citenamefont {Degroote}, \citenamefont {Heimonen}, \citenamefont {Kottmann}, \citenamefont {Menke} \emph {et~al.}}]{bharti2022noisy}%
  \BibitemOpen
  \bibfield  {author} {\bibinfo {author} {\bibfnamefont {K.}~\bibnamefont {Bharti}}, \bibinfo {author} {\bibfnamefont {A.}~\bibnamefont {Cervera-Lierta}}, \bibinfo {author} {\bibfnamefont {T.~H.}\ \bibnamefont {Kyaw}}, \bibinfo {author} {\bibfnamefont {T.}~\bibnamefont {Haug}}, \bibinfo {author} {\bibfnamefont {S.}~\bibnamefont {Alperin-Lea}}, \bibinfo {author} {\bibfnamefont {A.}~\bibnamefont {Anand}}, \bibinfo {author} {\bibfnamefont {M.}~\bibnamefont {Degroote}}, \bibinfo {author} {\bibfnamefont {H.}~\bibnamefont {Heimonen}}, \bibinfo {author} {\bibfnamefont {J.~S.}\ \bibnamefont {Kottmann}}, \bibinfo {author} {\bibfnamefont {T.}~\bibnamefont {Menke}}, \emph {et~al.},\ }\bibfield  {title} {\bibinfo {title} {Noisy intermediate-scale quantum algorithms},\ }\href@noop {} {\bibfield  {journal} {\bibinfo  {journal} {Reviews of Modern Physics}\ }\textbf {\bibinfo {volume} {94}},\ \bibinfo {pages} {015004} (\bibinfo {year} {2022})}\BibitemShut {NoStop}%
\bibitem [{\citenamefont {Preskill}(2018)}]{preskill2018quantum}%
  \BibitemOpen
  \bibfield  {author} {\bibinfo {author} {\bibfnamefont {J.}~\bibnamefont {Preskill}},\ }\bibfield  {title} {\bibinfo {title} {Quantum computing in the nisq era and beyond},\ }\href@noop {} {\bibfield  {journal} {\bibinfo  {journal} {Quantum}\ }\textbf {\bibinfo {volume} {2}},\ \bibinfo {pages} {79} (\bibinfo {year} {2018})}\BibitemShut {NoStop}%
\bibitem [{\citenamefont {Moll}\ \emph {et~al.}(2018)\citenamefont {Moll}, \citenamefont {Barkoutsos}, \citenamefont {Bishop}, \citenamefont {Chow}, \citenamefont {Cross}, \citenamefont {Egger}, \citenamefont {Filipp}, \citenamefont {Fuhrer}, \citenamefont {Gambetta}, \citenamefont {Ganzhorn} \emph {et~al.}}]{moll2018quantum}%
  \BibitemOpen
  \bibfield  {author} {\bibinfo {author} {\bibfnamefont {N.}~\bibnamefont {Moll}}, \bibinfo {author} {\bibfnamefont {P.}~\bibnamefont {Barkoutsos}}, \bibinfo {author} {\bibfnamefont {L.~S.}\ \bibnamefont {Bishop}}, \bibinfo {author} {\bibfnamefont {J.~M.}\ \bibnamefont {Chow}}, \bibinfo {author} {\bibfnamefont {A.}~\bibnamefont {Cross}}, \bibinfo {author} {\bibfnamefont {D.~J.}\ \bibnamefont {Egger}}, \bibinfo {author} {\bibfnamefont {S.}~\bibnamefont {Filipp}}, \bibinfo {author} {\bibfnamefont {A.}~\bibnamefont {Fuhrer}}, \bibinfo {author} {\bibfnamefont {J.~M.}\ \bibnamefont {Gambetta}}, \bibinfo {author} {\bibfnamefont {M.}~\bibnamefont {Ganzhorn}}, \emph {et~al.},\ }\bibfield  {title} {\bibinfo {title} {Quantum optimization using variational algorithms on near-term quantum devices},\ }\href@noop {} {\bibfield  {journal} {\bibinfo  {journal} {Quantum Science and Technology}\ }\textbf {\bibinfo {volume} {3}},\ \bibinfo {pages} {030503} (\bibinfo {year} {2018})}\BibitemShut {NoStop}%
\bibitem [{\citenamefont {Bravo-Prieto1}\ \emph {et~al.}(2023)\citenamefont {Bravo-Prieto1}, \citenamefont {LaRose}, \citenamefont {Cerezo}, \citenamefont {Subasi}, \citenamefont {Cincio1},\ and\ \citenamefont {Coles}}]{VQLSquantum}%
  \BibitemOpen
  \bibfield  {author} {\bibinfo {author} {\bibfnamefont {C.}~\bibnamefont {Bravo-Prieto1}}, \bibinfo {author} {\bibfnamefont {R.}~\bibnamefont {LaRose}}, \bibinfo {author} {\bibfnamefont {M.}~\bibnamefont {Cerezo}}, \bibinfo {author} {\bibfnamefont {Y.}~\bibnamefont {Subasi}}, \bibinfo {author} {\bibfnamefont {L.}~\bibnamefont {Cincio1}},\ and\ \bibinfo {author} {\bibfnamefont {P.~J.}\ \bibnamefont {Coles}},\ }\bibfield  {title} {\bibinfo {title} {{Variational Quantum Linear Solver}},\ }\href {https://doi.org/https://doi.org/10.22331/q-2023-11-22-1188} {\bibfield  {journal} {\bibinfo  {journal} {Quantum}\ }\textbf {\bibinfo {volume} {7}},\ \bibinfo {pages} {1188} (\bibinfo {year} {2023})}\BibitemShut {NoStop}%
\bibitem [{\citenamefont {Peruzzo}\ \emph {et~al.}(2014)\citenamefont {Peruzzo}, \citenamefont {McClean}, \citenamefont {Shadbolt}, \citenamefont {Yung}, \citenamefont {Zhou}, \citenamefont {Love}, \citenamefont {Aspuru-Guzik},\ and\ \citenamefont {O’brien}}]{peruzzo2014variational}%
  \BibitemOpen
  \bibfield  {author} {\bibinfo {author} {\bibfnamefont {A.}~\bibnamefont {Peruzzo}}, \bibinfo {author} {\bibfnamefont {J.}~\bibnamefont {McClean}}, \bibinfo {author} {\bibfnamefont {P.}~\bibnamefont {Shadbolt}}, \bibinfo {author} {\bibfnamefont {M.-H.}\ \bibnamefont {Yung}}, \bibinfo {author} {\bibfnamefont {X.-Q.}\ \bibnamefont {Zhou}}, \bibinfo {author} {\bibfnamefont {P.~J.}\ \bibnamefont {Love}}, \bibinfo {author} {\bibfnamefont {A.}~\bibnamefont {Aspuru-Guzik}},\ and\ \bibinfo {author} {\bibfnamefont {J.~L.}\ \bibnamefont {O’brien}},\ }\bibfield  {title} {\bibinfo {title} {A variational eigenvalue solver on a photonic quantum processor},\ }\href@noop {} {\bibfield  {journal} {\bibinfo  {journal} {Nature communications}\ }\textbf {\bibinfo {volume} {5}},\ \bibinfo {pages} {1} (\bibinfo {year} {2014})}\BibitemShut {NoStop}%
\bibitem [{\citenamefont {McClean}\ \emph {et~al.}(2016)\citenamefont {McClean}, \citenamefont {Romero}, \citenamefont {Babbush},\ and\ \citenamefont {Aspuru-Guzik}}]{mcclean2016theory}%
  \BibitemOpen
  \bibfield  {author} {\bibinfo {author} {\bibfnamefont {J.~R.}\ \bibnamefont {McClean}}, \bibinfo {author} {\bibfnamefont {J.}~\bibnamefont {Romero}}, \bibinfo {author} {\bibfnamefont {R.}~\bibnamefont {Babbush}},\ and\ \bibinfo {author} {\bibfnamefont {A.}~\bibnamefont {Aspuru-Guzik}},\ }\bibfield  {title} {\bibinfo {title} {The theory of variational hybrid quantum-classical algorithms},\ }\href@noop {} {\bibfield  {journal} {\bibinfo  {journal} {New Journal of Physics}\ }\textbf {\bibinfo {volume} {18}},\ \bibinfo {pages} {023023} (\bibinfo {year} {2016})}\BibitemShut {NoStop}%
\bibitem [{\citenamefont {Peng}\ \emph {et~al.}(2020)\citenamefont {Peng}, \citenamefont {Harrow}, \citenamefont {Ozols},\ and\ \citenamefont {Wu}}]{peng2020simulating}%
  \BibitemOpen
  \bibfield  {author} {\bibinfo {author} {\bibfnamefont {T.}~\bibnamefont {Peng}}, \bibinfo {author} {\bibfnamefont {A.~W.}\ \bibnamefont {Harrow}}, \bibinfo {author} {\bibfnamefont {M.}~\bibnamefont {Ozols}},\ and\ \bibinfo {author} {\bibfnamefont {X.}~\bibnamefont {Wu}},\ }\bibfield  {title} {\bibinfo {title} {Simulating large quantum circuits on a small quantum computer},\ }\href@noop {} {\bibfield  {journal} {\bibinfo  {journal} {Physical Review Letters}\ }\textbf {\bibinfo {volume} {125}},\ \bibinfo {pages} {150504} (\bibinfo {year} {2020})}\BibitemShut {NoStop}%
\bibitem [{\citenamefont {Lyu}\ \emph {et~al.}(2020)\citenamefont {Lyu}, \citenamefont {Montenegro},\ and\ \citenamefont {Bayat}}]{lyu2020accelerated}%
  \BibitemOpen
  \bibfield  {author} {\bibinfo {author} {\bibfnamefont {C.}~\bibnamefont {Lyu}}, \bibinfo {author} {\bibfnamefont {V.}~\bibnamefont {Montenegro}},\ and\ \bibinfo {author} {\bibfnamefont {A.}~\bibnamefont {Bayat}},\ }\bibfield  {title} {\bibinfo {title} {Accelerated variational algorithms for digital quantum simulation of many-body ground states},\ }\href@noop {} {\bibfield  {journal} {\bibinfo  {journal} {Quantum}\ }\textbf {\bibinfo {volume} {4}},\ \bibinfo {pages} {324} (\bibinfo {year} {2020})}\BibitemShut {NoStop}%
\bibitem [{\citenamefont {Tilly}\ \emph {et~al.}(2021)\citenamefont {Tilly}, \citenamefont {Chen}, \citenamefont {Cao}, \citenamefont {Picozzi}, \citenamefont {Setia}, \citenamefont {Li}, \citenamefont {Grant}, \citenamefont {Wossnig}, \citenamefont {Rungger}, \citenamefont {Booth} \emph {et~al.}}]{tilly2021variational}%
  \BibitemOpen
  \bibfield  {author} {\bibinfo {author} {\bibfnamefont {J.}~\bibnamefont {Tilly}}, \bibinfo {author} {\bibfnamefont {H.}~\bibnamefont {Chen}}, \bibinfo {author} {\bibfnamefont {S.}~\bibnamefont {Cao}}, \bibinfo {author} {\bibfnamefont {D.}~\bibnamefont {Picozzi}}, \bibinfo {author} {\bibfnamefont {K.}~\bibnamefont {Setia}}, \bibinfo {author} {\bibfnamefont {Y.}~\bibnamefont {Li}}, \bibinfo {author} {\bibfnamefont {E.}~\bibnamefont {Grant}}, \bibinfo {author} {\bibfnamefont {L.}~\bibnamefont {Wossnig}}, \bibinfo {author} {\bibfnamefont {I.}~\bibnamefont {Rungger}}, \bibinfo {author} {\bibfnamefont {G.~H.}\ \bibnamefont {Booth}}, \emph {et~al.},\ }\bibfield  {title} {\bibinfo {title} {The variational quantum eigensolver: a review of methods and best practices},\ }\href@noop {} {\bibfield  {journal} {\bibinfo  {journal} {arXiv preprint arXiv:2111.05176}\ } (\bibinfo {year} {2021})}\BibitemShut {NoStop}%
\bibitem [{\citenamefont {Huang}\ \emph {et~al.}(2022)\citenamefont {Huang}, \citenamefont {Li}, \citenamefont {Hou}, \citenamefont {Wu}, \citenamefont {Yung}, \citenamefont {Bayat},\ and\ \citenamefont {Wang}}]{huang2022robust}%
  \BibitemOpen
  \bibfield  {author} {\bibinfo {author} {\bibfnamefont {Y.}~\bibnamefont {Huang}}, \bibinfo {author} {\bibfnamefont {Q.}~\bibnamefont {Li}}, \bibinfo {author} {\bibfnamefont {X.}~\bibnamefont {Hou}}, \bibinfo {author} {\bibfnamefont {R.}~\bibnamefont {Wu}}, \bibinfo {author} {\bibfnamefont {M.-H.}\ \bibnamefont {Yung}}, \bibinfo {author} {\bibfnamefont {A.}~\bibnamefont {Bayat}},\ and\ \bibinfo {author} {\bibfnamefont {X.}~\bibnamefont {Wang}},\ }\bibfield  {title} {\bibinfo {title} {Robust resource-efficient quantum variational ansatz through an evolutionary algorithm},\ }\href@noop {} {\bibfield  {journal} {\bibinfo  {journal} {Physical Review A}\ }\textbf {\bibinfo {volume} {105}},\ \bibinfo {pages} {052414} (\bibinfo {year} {2022})}\BibitemShut {NoStop}%
\bibitem [{\citenamefont {Fedorov}\ \emph {et~al.}(2022)\citenamefont {Fedorov}, \citenamefont {Peng}, \citenamefont {Govind},\ and\ \citenamefont {Alexeev}}]{VQEFedorov}%
  \BibitemOpen
  \bibfield  {author} {\bibinfo {author} {\bibfnamefont {D.~A.}\ \bibnamefont {Fedorov}}, \bibinfo {author} {\bibfnamefont {B.}~\bibnamefont {Peng}}, \bibinfo {author} {\bibfnamefont {N.}~\bibnamefont {Govind}},\ and\ \bibinfo {author} {\bibfnamefont {Y.}~\bibnamefont {Alexeev}},\ }\bibfield  {title} {\bibinfo {title} {{VQE method: a short survey and recent developments}},\ }\href {https://doi.org/https://doi.org/10.1186/s41313-021-00032-6} {\bibfield  {journal} {\bibinfo  {journal} {Materials Theory}\ }\textbf {\bibinfo {volume} {6}},\ \bibinfo {pages} {2} (\bibinfo {year} {2022})}\BibitemShut {NoStop}%
\bibitem [{\citenamefont {McArdle}\ \emph {et~al.}(2019)\citenamefont {McArdle}, \citenamefont {Jones}, \citenamefont {Endo}, \citenamefont {Li}, \citenamefont {Benjamin},\ and\ \citenamefont {Yuan}}]{mcardle2019variational}%
  \BibitemOpen
  \bibfield  {author} {\bibinfo {author} {\bibfnamefont {S.}~\bibnamefont {McArdle}}, \bibinfo {author} {\bibfnamefont {T.}~\bibnamefont {Jones}}, \bibinfo {author} {\bibfnamefont {S.}~\bibnamefont {Endo}}, \bibinfo {author} {\bibfnamefont {Y.}~\bibnamefont {Li}}, \bibinfo {author} {\bibfnamefont {S.~C.}\ \bibnamefont {Benjamin}},\ and\ \bibinfo {author} {\bibfnamefont {X.}~\bibnamefont {Yuan}},\ }\bibfield  {title} {\bibinfo {title} {Variational ansatz-based quantum simulation of imaginary time evolution},\ }\href@noop {} {\bibfield  {journal} {\bibinfo  {journal} {npj Quantum Information}\ }\textbf {\bibinfo {volume} {5}},\ \bibinfo {pages} {1} (\bibinfo {year} {2019})}\BibitemShut {NoStop}%
\bibitem [{\citenamefont {Farhi}\ \emph {et~al.}(2014)\citenamefont {Farhi}, \citenamefont {Goldstone},\ and\ \citenamefont {Gutmann}}]{farhi2014quantum}%
  \BibitemOpen
  \bibfield  {author} {\bibinfo {author} {\bibfnamefont {E.}~\bibnamefont {Farhi}}, \bibinfo {author} {\bibfnamefont {J.}~\bibnamefont {Goldstone}},\ and\ \bibinfo {author} {\bibfnamefont {S.}~\bibnamefont {Gutmann}},\ }\bibfield  {title} {\bibinfo {title} {A quantum approximate optimization algorithm},\ }\href@noop {} {\bibfield  {journal} {\bibinfo  {journal} {arXiv preprint arXiv:1411.4028}\ } (\bibinfo {year} {2014})}\BibitemShut {NoStop}%
\bibitem [{\citenamefont {Kandala}\ \emph {et~al.}(2017)\citenamefont {Kandala}, \citenamefont {Mezzacapo}, \citenamefont {Temme}, \citenamefont {Takita}, \citenamefont {Brink}, \citenamefont {Chow},\ and\ \citenamefont {Gambetta}}]{kandala2017hardware}%
  \BibitemOpen
  \bibfield  {author} {\bibinfo {author} {\bibfnamefont {A.}~\bibnamefont {Kandala}}, \bibinfo {author} {\bibfnamefont {A.}~\bibnamefont {Mezzacapo}}, \bibinfo {author} {\bibfnamefont {K.}~\bibnamefont {Temme}}, \bibinfo {author} {\bibfnamefont {M.}~\bibnamefont {Takita}}, \bibinfo {author} {\bibfnamefont {M.}~\bibnamefont {Brink}}, \bibinfo {author} {\bibfnamefont {J.~M.}\ \bibnamefont {Chow}},\ and\ \bibinfo {author} {\bibfnamefont {J.~M.}\ \bibnamefont {Gambetta}},\ }\bibfield  {title} {\bibinfo {title} {Hardware-efficient variational quantum eigensolver for small molecules and quantum magnets},\ }\href@noop {} {\bibfield  {journal} {\bibinfo  {journal} {Nature}\ }\textbf {\bibinfo {volume} {549}},\ \bibinfo {pages} {242} (\bibinfo {year} {2017})}\BibitemShut {NoStop}%
\bibitem [{\citenamefont {Choquette}\ \emph {et~al.}(2021)\citenamefont {Choquette}, \citenamefont {Di~Paolo}, \citenamefont {Barkoutsos}, \citenamefont {S{\'e}n{\'e}chal}, \citenamefont {Tavernelli},\ and\ \citenamefont {Blais}}]{choquette2021quantum}%
  \BibitemOpen
  \bibfield  {author} {\bibinfo {author} {\bibfnamefont {A.}~\bibnamefont {Choquette}}, \bibinfo {author} {\bibfnamefont {A.}~\bibnamefont {Di~Paolo}}, \bibinfo {author} {\bibfnamefont {P.~K.}\ \bibnamefont {Barkoutsos}}, \bibinfo {author} {\bibfnamefont {D.}~\bibnamefont {S{\'e}n{\'e}chal}}, \bibinfo {author} {\bibfnamefont {I.}~\bibnamefont {Tavernelli}},\ and\ \bibinfo {author} {\bibfnamefont {A.}~\bibnamefont {Blais}},\ }\bibfield  {title} {\bibinfo {title} {Quantum-optimal-control-inspired ansatz for variational quantum algorithms},\ }\href@noop {} {\bibfield  {journal} {\bibinfo  {journal} {Physical Review Research}\ }\textbf {\bibinfo {volume} {3}},\ \bibinfo {pages} {023092} (\bibinfo {year} {2021})}\BibitemShut {NoStop}%
\bibitem [{\citenamefont {Lyu}\ \emph {et~al.}(2022)\citenamefont {Lyu}, \citenamefont {Xu}, \citenamefont {Yung},\ and\ \citenamefont {Bayat}}]{lyu2022symmetry}%
  \BibitemOpen
  \bibfield  {author} {\bibinfo {author} {\bibfnamefont {C.}~\bibnamefont {Lyu}}, \bibinfo {author} {\bibfnamefont {X.}~\bibnamefont {Xu}}, \bibinfo {author} {\bibfnamefont {M.}~\bibnamefont {Yung}},\ and\ \bibinfo {author} {\bibfnamefont {A.}~\bibnamefont {Bayat}},\ }\bibfield  {title} {\bibinfo {title} {Symmetry enhanced variational quantum eigensolver},\ }\href@noop {} {\bibfield  {journal} {\bibinfo  {journal} {arXiv preprint arXiv:2203.02444}\ } (\bibinfo {year} {2022})}\BibitemShut {NoStop}%
\bibitem [{\citenamefont {Wecker}\ \emph {et~al.}(2015)\citenamefont {Wecker}, \citenamefont {Hastings},\ and\ \citenamefont {Troyer}}]{wecker2015progress}%
  \BibitemOpen
  \bibfield  {author} {\bibinfo {author} {\bibfnamefont {D.}~\bibnamefont {Wecker}}, \bibinfo {author} {\bibfnamefont {M.~B.}\ \bibnamefont {Hastings}},\ and\ \bibinfo {author} {\bibfnamefont {M.}~\bibnamefont {Troyer}},\ }\bibfield  {title} {\bibinfo {title} {Progress towards practical quantum variational algorithms},\ }\href@noop {} {\bibfield  {journal} {\bibinfo  {journal} {Physical Review A}\ }\textbf {\bibinfo {volume} {92}},\ \bibinfo {pages} {042303} (\bibinfo {year} {2015})}\BibitemShut {NoStop}%
\bibitem [{\citenamefont {Wiersema}\ \emph {et~al.}(2020)\citenamefont {Wiersema}, \citenamefont {Zhou}, \citenamefont {de~Sereville}, \citenamefont {Carrasquilla}, \citenamefont {Kim},\ and\ \citenamefont {Yuen}}]{wiersema2020exploring}%
  \BibitemOpen
  \bibfield  {author} {\bibinfo {author} {\bibfnamefont {R.}~\bibnamefont {Wiersema}}, \bibinfo {author} {\bibfnamefont {C.}~\bibnamefont {Zhou}}, \bibinfo {author} {\bibfnamefont {Y.}~\bibnamefont {de~Sereville}}, \bibinfo {author} {\bibfnamefont {J.~F.}\ \bibnamefont {Carrasquilla}}, \bibinfo {author} {\bibfnamefont {Y.~B.}\ \bibnamefont {Kim}},\ and\ \bibinfo {author} {\bibfnamefont {H.}~\bibnamefont {Yuen}},\ }\bibfield  {title} {\bibinfo {title} {Exploring entanglement and optimization within the hamiltonian variational ansatz},\ }\href@noop {} {\bibfield  {journal} {\bibinfo  {journal} {PRX Quantum}\ }\textbf {\bibinfo {volume} {1}},\ \bibinfo {pages} {020319} (\bibinfo {year} {2020})}\BibitemShut {NoStop}%
\bibitem [{\citenamefont {Khatri}\ \emph {et~al.}(2019)\citenamefont {Khatri}, \citenamefont {LaRose}, \citenamefont {Poremba}, \citenamefont {Cincio}, \citenamefont {Sornborger},\ and\ \citenamefont {Coles}}]{khatri2019quantum}%
  \BibitemOpen
  \bibfield  {author} {\bibinfo {author} {\bibfnamefont {S.}~\bibnamefont {Khatri}}, \bibinfo {author} {\bibfnamefont {R.}~\bibnamefont {LaRose}}, \bibinfo {author} {\bibfnamefont {A.}~\bibnamefont {Poremba}}, \bibinfo {author} {\bibfnamefont {L.}~\bibnamefont {Cincio}}, \bibinfo {author} {\bibfnamefont {A.~T.}\ \bibnamefont {Sornborger}},\ and\ \bibinfo {author} {\bibfnamefont {P.~J.}\ \bibnamefont {Coles}},\ }\bibfield  {title} {\bibinfo {title} {Quantum-assisted quantum compiling},\ }\href@noop {} {\bibfield  {journal} {\bibinfo  {journal} {Quantum}\ }\textbf {\bibinfo {volume} {3}},\ \bibinfo {pages} {140} (\bibinfo {year} {2019})}\BibitemShut {NoStop}%
\bibitem [{\citenamefont {Duchi}\ \emph {et~al.}(2011)\citenamefont {Duchi}, \citenamefont {Hazan},\ and\ \citenamefont {Singer}}]{duchi2011adaptive}%
  \BibitemOpen
  \bibfield  {author} {\bibinfo {author} {\bibfnamefont {J.}~\bibnamefont {Duchi}}, \bibinfo {author} {\bibfnamefont {E.}~\bibnamefont {Hazan}},\ and\ \bibinfo {author} {\bibfnamefont {Y.}~\bibnamefont {Singer}},\ }\bibfield  {title} {\bibinfo {title} {Adaptive subgradient methods for online learning and stochastic optimization.},\ }\href@noop {} {\bibfield  {journal} {\bibinfo  {journal} {Journal of machine learning research}\ }\textbf {\bibinfo {volume} {12}} (\bibinfo {year} {2011})}\BibitemShut {NoStop}%
\bibitem [{\citenamefont {Bergholm}\ \emph {et~al.}(2018)\citenamefont {Bergholm}, \citenamefont {Izaac}, \citenamefont {Schuld}, \citenamefont {Gogolin}, \citenamefont {Ahmed}, \citenamefont {Ajith}, \citenamefont {Alam}, \citenamefont {Alonso-Linaje}, \citenamefont {AkashNarayanan}, \citenamefont {Asadi} \emph {et~al.}}]{bergholm2018pennylane}%
  \BibitemOpen
  \bibfield  {author} {\bibinfo {author} {\bibfnamefont {V.}~\bibnamefont {Bergholm}}, \bibinfo {author} {\bibfnamefont {J.}~\bibnamefont {Izaac}}, \bibinfo {author} {\bibfnamefont {M.}~\bibnamefont {Schuld}}, \bibinfo {author} {\bibfnamefont {C.}~\bibnamefont {Gogolin}}, \bibinfo {author} {\bibfnamefont {S.}~\bibnamefont {Ahmed}}, \bibinfo {author} {\bibfnamefont {V.}~\bibnamefont {Ajith}}, \bibinfo {author} {\bibfnamefont {M.~S.}\ \bibnamefont {Alam}}, \bibinfo {author} {\bibfnamefont {G.}~\bibnamefont {Alonso-Linaje}}, \bibinfo {author} {\bibfnamefont {B.}~\bibnamefont {AkashNarayanan}}, \bibinfo {author} {\bibfnamefont {A.}~\bibnamefont {Asadi}}, \emph {et~al.},\ }\bibfield  {title} {\bibinfo {title} {Pennylane: Automatic differentiation of hybrid quantum-classical computations},\ }\href@noop {} {\bibfield  {journal} {\bibinfo  {journal} {arXiv preprint arXiv:1811.04968}\ } (\bibinfo {year} {2018})}\BibitemShut {NoStop}%
\bibitem [{\citenamefont {Berry}\ \emph {et~al.}(2007)\citenamefont {Berry}, \citenamefont {Ahokas}, \citenamefont {Cleve},\ and\ \citenamefont {Sanders}}]{berry}%
  \BibitemOpen
  \bibfield  {author} {\bibinfo {author} {\bibfnamefont {D.~W.}\ \bibnamefont {Berry}}, \bibinfo {author} {\bibfnamefont {G.}~\bibnamefont {Ahokas}}, \bibinfo {author} {\bibfnamefont {R.}~\bibnamefont {Cleve}},\ and\ \bibinfo {author} {\bibfnamefont {B.~C.}\ \bibnamefont {Sanders}},\ }\bibfield  {title} {\bibinfo {title} {Efficient quantum algorithms for simulating sparse hamiltonians},\ }\href@noop {} {\bibfield  {journal} {\bibinfo  {journal} {Communications in Mathematical Physics}\ }\textbf {\bibinfo {volume} {270}},\ \bibinfo {pages} {359} (\bibinfo {year} {2007})}\BibitemShut {NoStop}%
\bibitem [{\citenamefont {Bultink}\ \emph {et~al.}(2020)\citenamefont {Bultink}, \citenamefont {O’Brien}, \citenamefont {Vollmer}, \citenamefont {Muthusubramanian}, \citenamefont {Beekman}, \citenamefont {Rol}, \citenamefont {Fu}, \citenamefont {Tarasinski}, \citenamefont {Ostroukh}, \citenamefont {Varbanov} \emph {et~al.}}]{bultink2020protecting}%
  \BibitemOpen
  \bibfield  {author} {\bibinfo {author} {\bibfnamefont {C.}~\bibnamefont {Bultink}}, \bibinfo {author} {\bibfnamefont {T.}~\bibnamefont {O’Brien}}, \bibinfo {author} {\bibfnamefont {R.}~\bibnamefont {Vollmer}}, \bibinfo {author} {\bibfnamefont {N.}~\bibnamefont {Muthusubramanian}}, \bibinfo {author} {\bibfnamefont {M.}~\bibnamefont {Beekman}}, \bibinfo {author} {\bibfnamefont {M.}~\bibnamefont {Rol}}, \bibinfo {author} {\bibfnamefont {X.}~\bibnamefont {Fu}}, \bibinfo {author} {\bibfnamefont {B.}~\bibnamefont {Tarasinski}}, \bibinfo {author} {\bibfnamefont {V.}~\bibnamefont {Ostroukh}}, \bibinfo {author} {\bibfnamefont {B.}~\bibnamefont {Varbanov}}, \emph {et~al.},\ }\bibfield  {title} {\bibinfo {title} {Protecting quantum entanglement from leakage and qubit errors via repetitive parity measurements},\ }\href@noop {} {\bibfield  {journal} {\bibinfo  {journal} {Science advances}\ }\textbf {\bibinfo {volume} {6}},\ \bibinfo {pages} {eaay3050} (\bibinfo {year} {2020})}\BibitemShut {NoStop}%
\bibitem [{\citenamefont {Roffe}\ \emph {et~al.}(2018)\citenamefont {Roffe}, \citenamefont {Headley}, \citenamefont {Chancellor}, \citenamefont {Horsman},\ and\ \citenamefont {Kendon}}]{roffe2018protecting}%
  \BibitemOpen
  \bibfield  {author} {\bibinfo {author} {\bibfnamefont {J.}~\bibnamefont {Roffe}}, \bibinfo {author} {\bibfnamefont {D.}~\bibnamefont {Headley}}, \bibinfo {author} {\bibfnamefont {N.}~\bibnamefont {Chancellor}}, \bibinfo {author} {\bibfnamefont {D.}~\bibnamefont {Horsman}},\ and\ \bibinfo {author} {\bibfnamefont {V.}~\bibnamefont {Kendon}},\ }\bibfield  {title} {\bibinfo {title} {Protecting quantum memories using coherent parity check codes},\ }\href@noop {} {\bibfield  {journal} {\bibinfo  {journal} {Quantum Science and Technology}\ }\textbf {\bibinfo {volume} {3}},\ \bibinfo {pages} {035010} (\bibinfo {year} {2018})}\BibitemShut {NoStop}%
\bibitem [{\citenamefont {McClean}\ \emph {et~al.}(2018)\citenamefont {McClean}, \citenamefont {Boixo}, \citenamefont {Smelyanskiy}, \citenamefont {Babbush},\ and\ \citenamefont {Neven}}]{mcclean2018barren}%
  \BibitemOpen
  \bibfield  {author} {\bibinfo {author} {\bibfnamefont {J.~R.}\ \bibnamefont {McClean}}, \bibinfo {author} {\bibfnamefont {S.}~\bibnamefont {Boixo}}, \bibinfo {author} {\bibfnamefont {V.~N.}\ \bibnamefont {Smelyanskiy}}, \bibinfo {author} {\bibfnamefont {R.}~\bibnamefont {Babbush}},\ and\ \bibinfo {author} {\bibfnamefont {H.}~\bibnamefont {Neven}},\ }\bibfield  {title} {\bibinfo {title} {Barren plateaus in quantum neural network training landscapes},\ }\href@noop {} {\bibfield  {journal} {\bibinfo  {journal} {Nature communications}\ }\textbf {\bibinfo {volume} {9}},\ \bibinfo {pages} {1} (\bibinfo {year} {2018})}\BibitemShut {NoStop}%
\bibitem [{\citenamefont {Cerezo}\ \emph {et~al.}(2021{\natexlab{b}})\citenamefont {Cerezo}, \citenamefont {Sone}, \citenamefont {Volkoff}, \citenamefont {Cincio},\ and\ \citenamefont {Coles}}]{cerezo2021cost}%
  \BibitemOpen
  \bibfield  {author} {\bibinfo {author} {\bibfnamefont {M.}~\bibnamefont {Cerezo}}, \bibinfo {author} {\bibfnamefont {A.}~\bibnamefont {Sone}}, \bibinfo {author} {\bibfnamefont {T.}~\bibnamefont {Volkoff}}, \bibinfo {author} {\bibfnamefont {L.}~\bibnamefont {Cincio}},\ and\ \bibinfo {author} {\bibfnamefont {P.~J.}\ \bibnamefont {Coles}},\ }\bibfield  {title} {\bibinfo {title} {Cost function dependent barren plateaus in shallow parametrized quantum circuits},\ }\href@noop {} {\bibfield  {journal} {\bibinfo  {journal} {Nature communications}\ }\textbf {\bibinfo {volume} {12}},\ \bibinfo {pages} {1} (\bibinfo {year} {2021}{\natexlab{b}})}\BibitemShut {NoStop}%
\end{thebibliography}%
\newpage
\appendix
\end{document}